# 200mm Optical synthetic aperture imaging over 120 meters distance via Macroscopic Fourier ptychography

**QI ZHANG**[1,2,3†], **YURAN LU**[4†], **YINGHUI GUO**[1,2,3,5,6], **YINGJIE SHANG**[1,2,3,5], **MINGBO PU**[1,2,3,5], **YULONG FAN**[1,2,3], **RUI ZHOU**[4], **XIAOYIN LI**[1,2,3], **AN PAN**[7], **FEI ZHANG**[1,2,3], **MINGFENG XU**[1,2,3], **XIANGANG LUO**[1,2,3,5*]

[1]*National Key Laboratory of Optical Field Manipulation Science and Technology, Institute of Optics and Electronics, Chinese Academy of Sciences, Chengdu 610209, P. R. China*
[2]*State Key Laboratory of Optical Technologies on Nano-Fabrication and Micro-Engineering, Institute of Optics and Electronics, Chinese Academy of Sciences, Chengdu 610209, P. R. China*
[3]*Research Center on Vector Optical Fields, Institute of Optics and Electronics, Chinese Academy of Sciences, Chengdu 610209, P. R. China*
[4]*Tianfu Xinglong Lake Laboratory, Chengdu 610299, P. R. China;*
[5]*College of Materials Sciences and Opto-Electronic Technology, University of Chinese Academy of Sciences, Beijing 100049, P. R. China;*
[6]*Sichuan Provincial Engineering Research Center of Digital Materials, Chengdu 610299, P. R. China;*
[7]*State Key Laboratory of Transient Optics and Photonics, Xi'an Institute of Optics and Precision Mechanics, Chinese Academy of Sciences, Xi'an 710119, P. R. China.*
*† These authors contribute equally: Qi Zhang, Yuran Lu.*
*\*E-mail: lxg@ioe.ac.cn*

**Abstract**: Fourier ptychography (FP) imaging, drawing on the idea of synthetic aperture, has been demonstrated as a potential approach for remote sub-diffraction-limited imaging. Nevertheless, the farthest imaging distance is still limited around 10 m even though there has been a significant improvement in macroscopic FP. The most severely issue in increasing the imaging distance is field of view (FoV) limitation caused by far-field condition for diffraction. Here, we propose to modify the Fourier far-field condition for rough reflective objects, aiming to overcome the small FoV limitation by using a divergent beam to illuminate objects. A joint optimization of pupil function and target image is utilized to attain the aberration-free image while estimating the pupil function simultaneously. Benefiting from the optimized reconstruction algorithm which effectively expands the camera's effective aperture, we experimentally implement several FP systems suited for imaging distance of 12 m, 65 m and 120m with the maximum synthetic aperture of 200 mm. The maximum synthetic aperture is thus improved by more than one order of magnitude of the state-of-the-art works from the furthest distance, with an over fourfold improvement in the resolution compare to single aperture. Our findings demonstrate significant potential for advancing the field of macroscopic FP, propelling it into a new stage of development.

## 1. Introduction

Past decades have witnessed fast-growing sub-diffraction imaging techniques. Although near-field techniques such as superlens [1–3], scanning near-field optical microscopy [4,5], microsphere-assisted imaging [6], can achieve deep sub-wavelength resolution, their applications are quite limited to the flat objects and their short working distances on the order of tens of nanometers. To date, high-resolution imaging of distant targets is still challenging in applications such as surveillance, remote sensing, and astronomy. The fundamental limitation of imaging resolution in these scenarios is well known, defined by the diffraction limit of $1.22\lambda L/D$, where $\lambda$ denotes the wavelength, $L$ is the object distance, and $D$ is the diameter of the aperture. Increasing the aperture of a lens is a usual choice to improve the imaging resolution, nevertheless it incurs proportionally larger aberrations. The addition of extra optical elements can correct these aberrations, but significantly increase the volume and weight of the system [7]. Alternatively, synthetic aperture is an effective method to achieve large equivalent aperture that can be orders of magnitude larger than its physical aperture. Synthetic aperture imaging has been broadly exploited in microwave technologies such as large radio telescopes, [8] synthetic aperture radar (SAR) [9], and very-long-baseline interferometry (VLBI) [10]. This success is attributed to the direct measurability of both the intensity and phase of the incident microwaves in the radio frequency domain. Inspired by these innovative concepts, optical counterparts have been proposed, including distributed aperture synthesis [11], synthetic aperture lidar [12,13], and Segmented Planar Imaging Detector for Electro-optical Reconnaissance (SPIDER) [14,15]. Nevertheless, all these techniques face an inevitable challenge in precise phase matching among different optical paths, given that conventional CCD cameras cannot directly resolve the phase of incident waves.

Computational optical imaging is a revolutionary technology capable of recovering the lost information through post-processing using incomplete physical measurements. Technically, computational imaging breaks through the point-to-point mapping used in conventional imaging systems, thus allowing for the enhancement of the system's performance without demand for a larger optical system. The relevant applications range from high security-level optical cryptography [16–18], thin metalens based full-color imaging [19,20] to polarization and spectral imaging/detection, [21–25] scattering medium imaging [26,27], non-line-of-sight imaging [28–30], among others [31–33]. Herein, we intentionally focus on the computational imaging to recover image details well below the diffraction limit of very distant objects.

Fourier ptychography (FP), as one typical kind of computational imaging techniques, combines the phase retrieval process with synthetic aperture technique, enabling sub-diffraction imaging without stringent requirement of phase matching [34,35]. Microscopic FP technique acquires the Fourier components dependent intensity images directly by an angle-resolved programmable LED array, or by aperture-scanning at the Fourier plane [34]. Advanced FP phase retrieval algorithms [36–43] are then applied to reconstruct images from the complex expanded Fourier spectrum, resulting in a sub-diffraction imaging resolution much better than the resolution limit imposed by the objective. Recently, there is growing interest in extending FP imaging to macroscopic realm [44–49], especially via reflective framework. Using a photographic lens as a Fourier transformer at its focal plane, reflective-type FP acquires image sequence by simply scanning the camera to different positions perpendicular to the optical axis. In principle, the maximum achievable resolution of such a mechanism is determined by the camera's traveling range, provided the sampling rate of the camera is sufficient.

Quite recently, many progresses have been made in macroscopic FP, make FP a much powerful methods for sub-diffraction imaging. However, the maximum experimental distance of reflective FP remains around 10 m [50,51]. In order to satisfy the far-field condition ($z \geq 2D^2/\lambda$) in the macroscopic FP, the FoV is usually limited. Furthermore, to avoid the aberration caused by the photographic lens, a small effective aperture (~2.5 mm) is typically utilized (while the full aperture is much larger), leads to relatively low synthetic aperture [45,46]. These issues limit the application of macroscopic FP. In this work, we show that the far-field condition for an optically rough object can be rewritten as $z \geq 2DR_c/\lambda$, where $R_c$ means correlation radius of the scattering object to be imaged, making FoV of FP imaging enlarged from the limitation of illumination system. Based on this principle, we devise and experimentally demonstrate several FP systems with divergent illuminated beam illumination. A joint optimization of pupil function and target image is utilized to eliminate the system's aberration, making full aperture of camera lens is practicable. Our results manifest a maximum imaging distance (65 m) and a maximum synthetic aperture (200 mm) improved by more than one order of magnitude of the state-of-the-art works with a fourfold improvement in imaging resolution.

## 2. Principle of reflective-type FP imaging

In reflective-type FP imaging, the ideal illumination source and the ideal imaging model are usually adopted as the forward propagation model. In this case, the light source is regarded as a monochromatic point source with no bandwidth, which is therefore completely coherent in time and spatial. After impinging on the surface of the object, a spatial modulation of amplitude and phase is imposed to the back-propagation waves before entering the diffraction-limited optical system. The intensity images are ultimately recorded by detectors without wavefront difference information. When assuming that the complex amplitude of the light modulated by the object surface is $U(\xi,\eta)$, after a propagation distance z, the back-reflected waves can be expressed as [52]:

$$U(x,y) = \frac{z}{j\lambda} \iint_\Sigma U(\xi,\eta) \frac{\exp(jkr_{01})}{r_{01}^2} d\xi d\eta \tag{1}$$

where $r_{01} = \sqrt{z^2 + (x-\xi)^2 + (y-\eta)^2}$ represent the distance between points in object plane at $(\xi,\eta,0)$ and observer plane at $(x,y,z)$, and $k$ is the wave vector. When the propagation distance $z$ meets the Fraunhofer far-field diffraction condition $z \gg k(\xi^2 + \eta^2)/2$, the light field distribution at the pupil of the imaging system could be simplified as:

$$U(x,y) = \frac{e^{jkz} e^{jk/2z}(x^2 + y^2)}{j\lambda z} \mathcal{F}_{\frac{1}{\lambda z}}\{U(\xi,\eta)\} \tag{2}$$

where $\mathcal{F}$ represents the two-dimensional Fourier transform, and $\frac{1}{\lambda z}$ is the unit length in the frequency domain. This light field distribution recorded by the restrict-apertured imaging system with no wavefront difference and vignetting, can be expressed as

$$P(\mathbf{x}-\mathbf{s}) = \begin{cases} 1 & \text{if } \|\mathbf{x}-\mathbf{s}\| < \frac{d}{2} \\ 0 & \text{otherwise} \end{cases} \tag{3}$$

where $d$ is the pupil aperture, $\mathbf{s}$ represents the center shift of the pupil. Then the optical field blocked by the aperture can be written as

$$U_P(\mathbf{x},\mathbf{s}) = U(\mathbf{x}) \cdot P(\mathbf{x}-\mathbf{s}) \tag{4}$$

When the light propagates from the pupil to the focal plane satisfying the Fraunhofer diffraction approximation condition, its field distribution can be directly described by the Fourier transform. Due to the lack of responsivity in the phase information, the intensity information recorded by the optical detector can be represented as

$$I(\mathbf{x},\mathbf{s}) \propto |\mathcal{F}[U_P(\mathbf{x},\mathbf{s})]|^2 \tag{5}$$

In the process of FP imaging, by moving the pupil position, that is, changing $P(\mathbf{s})$, the different intensity information can be recorded, and the light field information $U(\mathbf{x})$ at pupil plane can be recovered through the phase recovery iterative algorithm. This process equivalently increases the effective pupil size of the imaging system, which leads to the resolution enhancement according to the diffraction limited imaging theory $\theta_{min} \propto \lambda / D$.

Please note that there is a very strict condition in the above imaging model, namely the Fraunhofer (far-filed) diffraction approximation condition. In order to meet this condition, it is necessary to reduce the size of the object surface illuminated by a plane wave, or increase the object-image distance. In most cases, the object is illuminated with convergent beam to compensate the quadratic phase distribution required by the far-field condition, resulting in a very limited illuminated area. As a result, the size of illuminated area is normally half of the largest lens' aperture in the illuminated light path. Another way to address this problem is to add a fixed lens in front of the pupil of the imaging system as a Fourier lens, [50] whose size is much larger than the imaging system's aperture, degrading the FP imaging's advantage in achieving high resolution by using small aperture. By adding virtual quadratic phase distribution to objects, far-distance aperture synthesis can also be achieved even illumination with a divergent beam but the FoV was still limited. By considering the roughness of the object surface, the FoV can be extended due to the angle of reflected light from the target object get broader[51], It confirmed that FP can improve the resolution to the theoretical limit of the virtual synthetic aperture rather than simply suppressing the speckle. In this paper, we will discuss this model from the perspective of partial coherence.

In practical experiments, both the bandwidth of the laser source and the roughness of the object's surface will reduce the coherence of the back-reflected light field. As a result, the far-field conditions derived based on fully coherent light fields are too strict for the actual FP imaging[52, 53]. We can then resort to the far-field diffraction theory under partial coherence to calculate the complex light field at some distance $z$ away from the object plane using Fresnel integral of propagating mutual transmitted intensity $J_t$:

$$U(x,y) \approx \frac{1}{(\lambda z)^2} \iint_{-\infty}^{\infty} \iint_{-\infty}^{\infty} J_t(\xi_1, \eta_1; \xi_2, \eta_2) exp\left[-j\frac{2\pi}{\lambda}(r_2^{'} - r_1^{'})\right] d\xi_1 d\eta_1 d\xi_2 d\eta_2 \qquad (6)$$

where $r^{'}_{1,2}$ represent the distance between two planes. If we define $\Delta\xi = \xi_1 - \xi_2$, $\Delta\eta = \eta_1 - \eta_2$, $\bar{\xi} = (\xi_1 + \xi_2)/2$, $\bar{\eta} = (\eta_1 + \eta_2)/2$, then the field distribution at aperture plane of FP imaging could be rewritten as

$$U(x,y) \approx \frac{1}{(\lambda z)^2} \iint_{-\infty}^{\infty} \iint_{-\infty}^{\infty} J_t(\xi_1, \eta_1; \xi_2, \eta_2) exp\left[-j\frac{2\pi}{\lambda}(\bar{\xi}\Delta\xi + \bar{\eta}\Delta\eta - x\Delta\xi - y\Delta\eta)\right] d\xi_1 d\eta_1 d\xi_2 d\eta_2 \qquad (7)$$

The Fraunhofer far-field approximation condition in which satisfied Fourier transform is given by: $z \gg (\bar{\xi}\Delta\xi + \bar{\eta}\Delta\eta)/\lambda$. Here, the mutual transmitted intensity function $J_t$ is relative to the complex coherence factor $\mu(\Delta\xi, \Delta\eta)$, which is zero when two spots are not adjacent, that out of the coherence area. So in forward propagation model, the maximum value of $\bar{\xi}$, $\bar{\eta}$ is refer to full FoV, which is $D/2$; on the other hand, the maximum value for $\Delta\xi$, $\Delta\eta$ ensure $\mu(\Delta\xi, \Delta\eta)$ has a non-zero value is the correlation radius of object $R_c/2$. Consequently, the "far-field" condition reduces to: $z \gg DR_C/4\lambda$. It should be emphasized that the decrease of coherence is mainly spatial coherence, as a random spatial phase added on the illumination light at the surface of objects. The temporal coherence of light was still guaranteed by the illumination laser with sufficiently narrow linewidth. For spatial coherence, it has been demonstrated that the far-field condition with small correlation radius is effective in other applications such as imaging through turbid media and orbital-angular-momentum sorting [53,

54]. According to this model in FP imaging, less correlation radius ($R_c$) of object lead to larger FoV ($D$). We test this model with simulation (Supplementary Information section 1 and Figure S1) and based on this prerequisite, we conducted a series of experiments with divergent light source as discussed in the following context.

## 3. Results and discussion

In this section, we will present the experiment setup and results of reflective-type FP imaging at 12 m, and 65 m, respectively.

### 3.1 Reflective-type FP imaging at 12 m

The experimental setup constructed on the first platform is shown in Figure 1, whose optical path diagram is shown in Figure 1a and the physical setup shown in Figure 1b. A 532 nm fiber laser (Precilasers YFL-SSHG-532-10-CW) was used as light source, 100mW output power was applied in this experiment. Two lenses are used to expand the laser beam within an appropriate divergence to illuminate the target located 12.8 meters away on the second platform. In this experiment, the illuminated FoV is about Φ100 mm. The target of a resolution chart printed on a paper produces diffuse reflection for FP imaging. A CMOS image sensor with a pixel size of 2.2 um (MER2-507-23GM-P) was used to record the reflected signals. In front of the sensor situates a photographic lens with a focal length of 75 mm and aperture of 12 mm (f/6.75). A polarizer placed in front of the diaphragm is used to filter out the specular reflection component. The whole camera was installed on a XZ-axis translation stage. The translate stage has an one-way positioning accuracy less than 30 um, which is enough for the aperture diameter of our imaging lens. A grid of 21 × 21 images is captured to produce a synthetic aperture of 48 mm, 4 times of the lens' aperture. Adjacent positions of the camera are 1.8 mm apart to ensure 85% overlapping ratio between images in the Fourier domain.

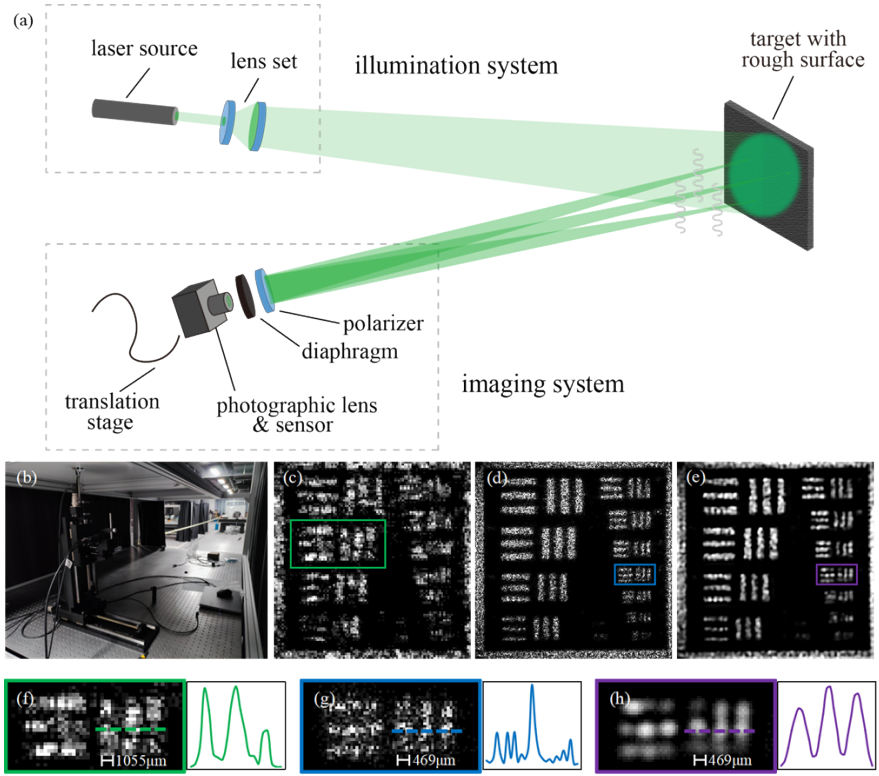

**Figure 1.** Experimental setup and results for reflective-type FP at 12.8m. a) A simplified optical path diagram of FP system. b) The target and the FP system are placed on different optical platforms and separated by 12.8 meters. c, d, e) The comparing result of single aperture image, synthetic aperture image, and denoised image of 12.8m FP and f, g, h), The corresponding zoom-in details of the detected images and the plot profile of the smallest distinguishable line pairs.

To reconstruct the image, we employed an optimized recover algorithm base on Gerchberg–Saxton (GS) algorithm for post-processing [46], allowing to recover the image captured by a camera with slight aberration. We describe this algorithm in detail in Methods section and discuss the influence of pupil function on the quality of recovered images in Supplementary Information section 2. The raw image captured by the single aperture is shown in Figure 1c, with zoom-in view of the red framed region and the corresponding intensity profile shown in Figure 1f. In comparison, the reconstructed FP image is shown in Figure 1d, with zoom-in view of the blue framed region and the corresponding intensity profile in Figure 1g. Clearly, the reconstructed image exhibits much better resolution (469 μm as compared to 1055 μm for raw image) with less noise and smaller speckles. To further constrain the speckles caused by coherent illumination, we adopted speckle suppression processing to reconstructed image after synthetic reconstruction, shown in Figure 1e and h. That make intensity profile of line pair can be quantitative analyzed while speckled images introduce many fluctuations. Here, we define the resolution according to the peak-valley contrast threshold over 0.5. The contrast is defined as

$$C = \frac{w-b}{w} \tag{8}$$

where $w$ is the peak value of the line-pair and $b$ is the valley value. In conclusion, a 2.3 times improvement in imaging resolution is achieved. The gap between theoretical improvement calculated by synthetic aperture and practical improvement might be caused by deficiency of sampling rate, while we used a large aperture ($F/6.75$) lens. Compared to a small aperture, large aperture lens acquired images with relatively insufficient sampling rate due to the optical resolution is relatively high.

### 3.2 Reflective-type FP imaging at 65 m

Furthermore, we extend the experimental scene to outdoors, performing FP imaging experiment between two buildings at a distance of 65 m, as shown in Fig. 2a. In this scenario, we replace the photographic lens with a telephoto lens with a focal length of 800 mm and an aperture diameter of 50 mm ($F/16$). The output power of laser was set to 2.9W, and the FoV was about Φ600 mm. We removed the polarizer in front of the lens due to the minor influence imposed by the specular reflection component in this experiment. A grid of 16 × 16 images was captured to produce a synthetic aperture of 200 mm, four times larger than the lens' physical aperture. The distance between adjacent positions of the camera is 10 mm to ensure an overlapping rate of 80%. We took a new resolution chart hanged over a grayscale photo of Solvay Conference as our new target, as shown in Fig. 2b. The resolution chart is resized to be optically resolvable in the low-resolution mode, as revealed in the left part of Figure 2c. The smallest distinguishable line pairs of the resolution chart along with the intensity profile in the low-resolution mode are shown in Fig. 2d, manifesting an imaging resolution of ~ 2 mm. Under this scheme, the photo of Solvay Conference is hardly to be recognized. In contrast, the FP imaging results revealed in the right part of Fig. 2c shows up a much clear image with less noise and speckles within the whole image area. A much better imaging resolution of 0.6 mm is achieved, as manifested by Fig. 2e. Under this scheme, almost all the features in the Photo of Solvay Conference can be distinguished, as demonstrated by the right part of Fig. 2c. A 3.3 times improvement in imaging resolution has been achieved.

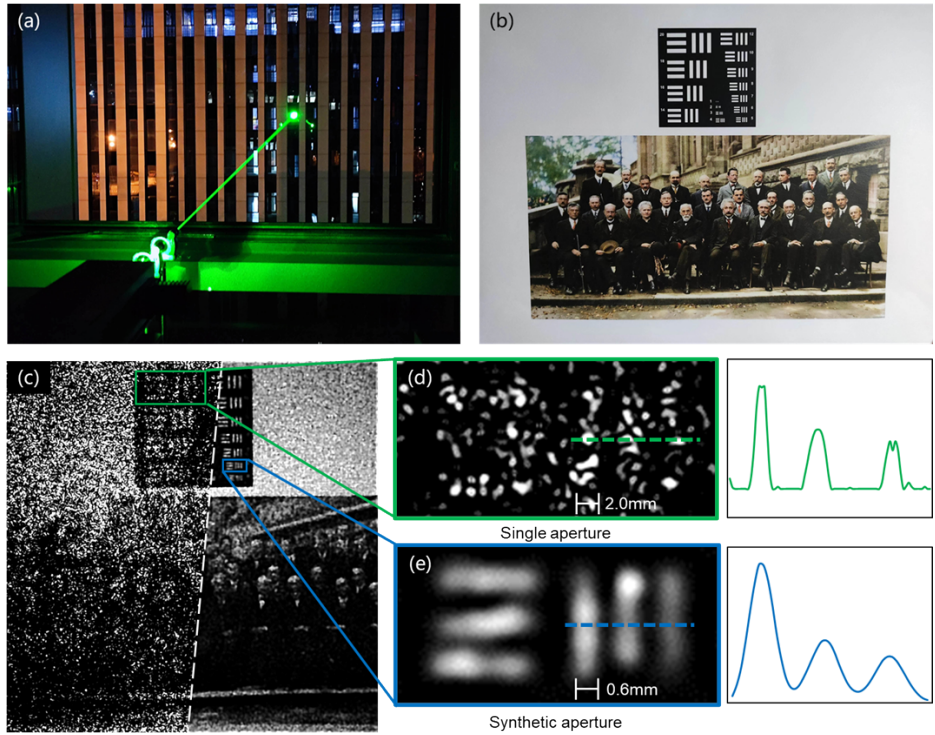

**Figure 2.** Experimental setup and results for 65m FP. a) Experimental scene of 65m FP. The distance between the target and the FP system is about 65 meters. b) The target used in 65m FP, consisting of binary image (upper part) and grayscale image (lower part). c) Results of resolution chart image and greyscale image. Left part shows the raw image recorded by single aperture, while right part shows the reconstructed result. d) and e) shows the enlarged picture of the smallest distinguishable line pairs in c) and the corresponding intensity profiles.

*3.3 Influence of intensity attenuation and phase aberration*

In order to explore the FP's performance at long distance, we performed FP imaging experiment in the short-wave infrared ray (SWIR) region because of its higher safety and turbulence resistance in active imaging than visible light. We chose a SWIR laser (Precilasers YFL-1064-10-CW) with wavelength of 1064 nm, the same 800 mm telephoto lens and another SWIR detector (Allied Vision Goldeye G-130) with pixel size of 5 μm to carry out this experiment. We replaced the target to a larger resolution chart to deal with the decrease in absolute resolution due to the increased wavelength. The distance between the target and the FP system is about 65 meters. A grid of 16 × 16 images was captured to produce a synthetic aperture of 200 mm, four times larger than the lens' aperture.

The result of SWIR FP at 65 m are shown in Figure 3a, manifesting a nearly four times resolution improvement. The reconstructed imaging result shown in Figure 3a2 and 3a3 exhibited a distinguishable line-pair at (-4, 2), while in image acquired by single aperture (Fig. 3a1) the minimum distinguished line-pair appeared on (-1, 2) with a line-width at 890 um, which is approximately 1.9 times of the diffraction limit (1.688 mm at the object plane). Considering the resolution reduction in coherent illumination imaging, FP imaging improves the resolution even more significantly (about 3.8 times of the coherent diffraction limit). Furthermore, we repeated the same experiment except the addition of a window between the telephoto lens and the target, introducing weak turbulence to the system. The results shown in Figure 3b demonstrate the robustness and stable functioning of SWIR FP system with the same resolution despite the weak turbulence and giant attenuation (~ 11 dB) imposed by the window, the latter of which is commonly regarded as a fatal issue in far-field imaging. Finally, we replaced the resolution chart with a badge with a symbol of sailboat as a target, the reconstructed result of FP image can unambiguously resolve more detail of the sailboat (middle panel of Fig. 3c) than the raw image (left panel of Fig. 3c). Moreover, the speckle suppression processing seems improve image quality more obvious in objects with dark feature in white background, shown in right panel of Figure 3c.

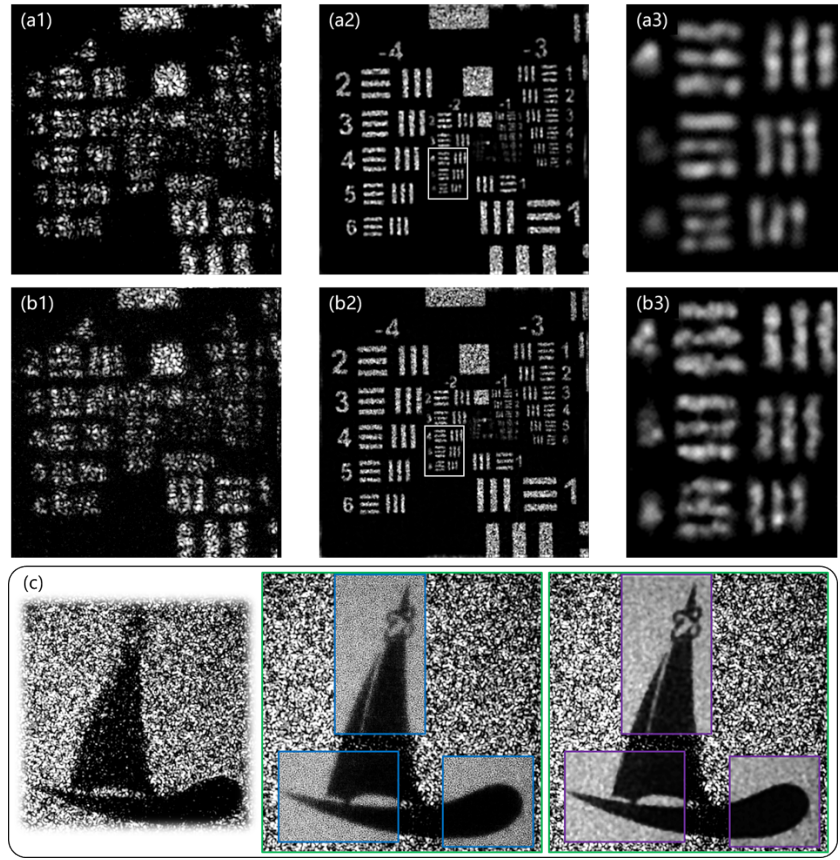

**Figure 3.** Results of 65m SWIR FP imaging. a) Imaging without a window in front of the lens. a1) single aperture raw image. a2) The reconstructed synthetic aperture image. a3) enlarged image of a2). b) Imaging with one single closed window in front of the lens. b1) single aperture raw image. b2) The reconstructed synthetic aperture image. b3) enlarged image of a2). c) Imaging of the printed pattern without window in between, with raw image shown on the left, the reconstructed image of FP imaging superposed with the raw image shown in the middle, and the denoised image shown on the right.

### *3.4 Reflective-type FP imaging of real-life scene at 120m*

In addition to the experiments on binary and grayscale target imaging presented above, we adopted FP imaging of a real-life scene in the night exclude the effect of sunlight. Using the same equipment of section 3.3, we performed the reflective-type FP imaging of a wire pole 120 meters away. The laser power is adjusted to 1.02 W. The exposure time of each image was set to 400 ms, which is short enough to restrain the blur effect caused by micro-vibration of the building or the pole. The experimental setup, the targeted wire pole in the daytime and the experimental scenario of the 120m FP imaging are shown in Figs. 4a, 4b. Screws on the pole (marked with yellow triangle) were far better resolved in the recovered image in comparison with the raw image, and the speckle suppression processing bring image higher contrast, as demonstrated in Figs. 4c-4e. It firmly demonstrates the effectiveness of FP imaging of real-life scenes with a work distance up to hundreds of meters. It's worth pointing out that the bright line in the center of both the raw image and the recovered images were caused by the specular reflection from the pole, resulting in the overexposure and unrecoverability in this area.

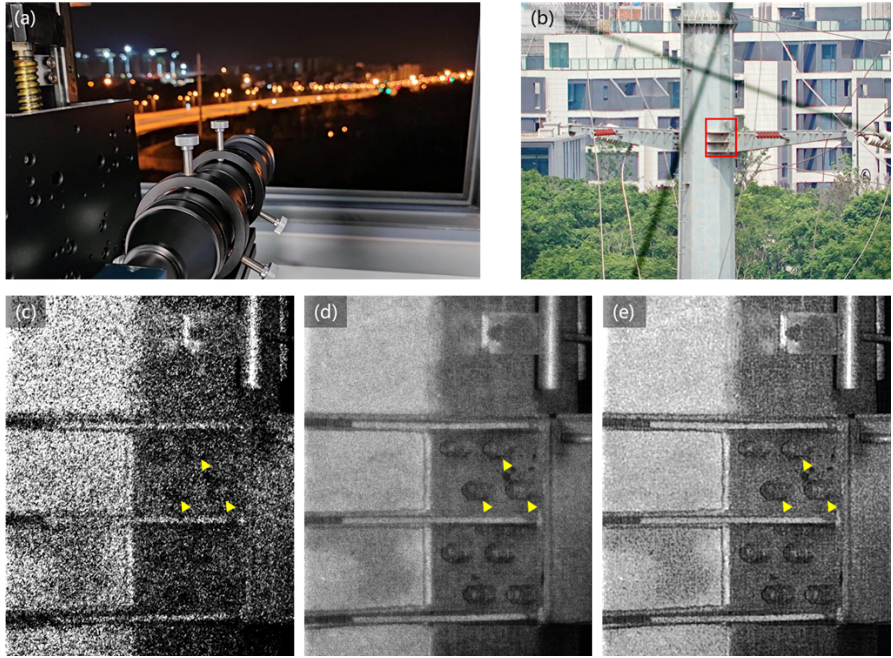

**Figure 4.** Imaging scene and results of 120m FP. a) Setup of FP imaging system and the practical imaging circumstances in evening. b) The wire pole target, the distance between the wire pole and the FP system is about 120 meters. c), d) and e) enlarged FP imaging results of red framed region in b). c) Single aperture result. d) Synthetic aperture result. e) Denoised result.

To additional improve the ability of long-range imaging in our system, we raise the illumination power from about 0.1W at 12m experiment to 2.9W and 1.02W at 65m and 120m experiment respectively (with different FoV and different object). According to our experiment the laser techniques at state of the art, the laser power is enough for 1km-level FP imaging. But imaging at greater distances will require more novel technology or model adjustments, such as using single-photon avalanche diode (SPAD) cameras [55,56] to improve detective sensitivity.

## 4. Conclusion and perspective

Reflective-type FP imaging methods have showcased its powerful capability in macroscopic super-resolution imaging. So far, they are only demonstrated in the laboratory condition, with the imaging distance on the order of 10 meters. We successfully extended the application scope of the FP imaging from laboratory condition to outdoor scene by introducing the "rough" far-field condition to FP imaging model. By using divergent light to illuminate objects, we achieved super-resolution imaging of objects located 12 m and 65 m away, with the maximum synthetic aperture of 200 mm. By optimizing the reconstruction method, we obtained a fourfold resolution improvement even with weak turbulence and giant optical attenuation. Our innovative methods hold the potential to drive the applications of FP imaging for practical applications across a broad field. Recently, flat telescopes composed of an all-dielectric metasurface doublet [57] or cascaded liquid crystal planar optical elements [58] have been proposed. Especially, the maximum aperture of liquid crystal planar optical elements have beyond 300 mm [59,60], such lightweight and planar optical elements could bring new possibilities for kilometer level FP imaging. Moreover, combination of multi-channel simultaneous acquisition (array camera) and deep learning to improve the image acquisition speed and process speed is also an important development trend of far-field FP imaging.

However, after crossing the 100-meter scale, FP imaging will encounter many engineering problems. The first is insufficient power. Active imaging with diffuse reflection at very long distances requires extremely high light sources. In order to avoid motion blur caused by mechanical jitter, our experiment time is also greatly extended at longer distances and larger aperture, and long-term database acquisition also makes FP more susceptible to atmospheric turbulence. Operating wavelength is switched to shortwave infrared to deal with atmospheric turbulence, but this is only useful when the atmospheric coherence length is large. When the atmospheric coherence length is low, which means the turbulence of imaging channel is strong, the temporal coherence of light will decrease severely. Due to that, the image sequences of single aperture will show smear and time-varying that will influence or even invalidate image reconstruction. The challenge of remote FP is severe.

## 5. Methods

At present, several recovery algorithms can be used for FP imaging. We adopted the alternate projection method based on GS algorithm and SAVI method for further optimization. At the beginning of the algorithm, we need to initialize the object spectrum and pupil function in Fourier domain. The algorithm aims at the iterative reconstruction of object spectrum $\Psi^i(u)$ and pupil function $P^i(u)$. The initial values of $\Psi^i(u)$ and $P^i(u)$ are given as

$$\Psi^i(u) = \mathcal{F}\left\{\frac{\sum_{n=1}^{N}\sqrt{I(x,c_n)}}{N}\right\} \tag{9}$$

$$P^i(u) = \begin{cases} 1 & if\ |u| < \frac{d}{2} \\ 0 & otherwise \end{cases} \tag{10}$$

where $I(x,c_n)$ is the low-resolution image taken at the $n$th position in $N$, $\Phi(u,c_n) = \Psi(u-c_n)P(u)$. That is, the image spectrum $\Phi(u,c_n)$ is the product of the object spectrum and the pupil function at the corresponding position.

Before the iteration cycle, the initial value of the object spectrum should be limited to avoid introducing unnecessary errors in subsequent iterations and causing algorithm divergence:

$$\Psi^i(u) = \Psi^i(u)\sum_{n=1}^{N}P^i(u+c_n) \tag{11}$$

where $\sum_{n=1}^{N}P^i(u+c_n)$ is the synthetic aperture formed by a series of pupil functions according to the reciprocal of their projection weights. The following part is the loop body of this algorithm. After the object spectrum is limited by pupil, the image spectrum is obtained:

$$\Phi^i(u,c_n) = \Psi^i(u-c_n)P^i(u) \tag{12}$$

The image spectrum propagates backward to obtain the field intensity distribution:

$$E^i(u,c_n) = \mathcal{F}^{-1}\{\Phi^i(u,c_n)\} \tag{13}$$

The phase of the field intensity distribution is preserved, and its intensity is replaced by the field intensity of the series of low-resolution images taken:

$$E^{i+1}(x,c_n) = \sqrt{I(x,c_n)} \frac{E^i(u,c_n)}{|E^i(u,c_n)|} \qquad (14)$$

The forward Fourier transform of the field intensity distribution allows to obtain the updated object spectrum:

$$\Phi^{i+1}(u,c_n) = \mathcal{F}\{E^{i+1}(x,c_n)\} \qquad (15)$$

The difference between the old and new image spectra is stacked with a certain weight and then the new object spectrum is calculated as:

$$\Psi^{i+1}(u) = \Psi^i(u) + \sum_{n=1}^{N} \frac{|P^i(u+c_n)|}{|P^i(u)|_{max}} \frac{[P^i(u+c_n)]^*}{|P^i(u+c_n)|^2 + \tau_1} \times [\Phi^{i+1}(u-c_n) - \Phi^i(u-c_n)] \qquad (16)$$

which is the key step of FP iteration–. The pupil function is updated in a similar way but with a strict limit to prevent the reconstruction process from diverging:

$$\Delta P^i(u) = \sum_{n=1}^{N} \frac{|\Psi^i(u-c_n)|}{|\Psi^i(u)|_{max}} \frac{[\Psi^i(u-c_n)]^*}{|\Psi^i(u-c_n)|^2 + \tau_2} \times [\Phi^{i+1}(u-c_n) - \Phi^i(u-c_n)]/N \qquad (17)$$

$$P^{i+1}(u) = P^i(u)*(1-\gamma) + \frac{\Delta P^i(u)}{|\Delta P^i(u)|_{max}} *\gamma \qquad (18)$$

The normalization in 10 is designed to prevent excessive local pupil changes resulting in pupil non-convergence. We can also multiply the pupil by $P^0(u)$ to suppress the extrinsic pupil error:

$$P^{i+1}(u) = P^{i+1}(u) \cdot P^0(u) \qquad (19)$$

The synthetic aperture spectrum $\Psi^{i_{max}}(u)$ can be obtained by repeating formulas 4 to 11, and the resultant high-resolution image can thus be obtained:

$$I_h(x) = [\mathcal{F}^{-1}\{\Psi^i(u)\}]^2 \qquad (20)$$

The complete recovery algorithm flow chart is shown in (Figure 5).

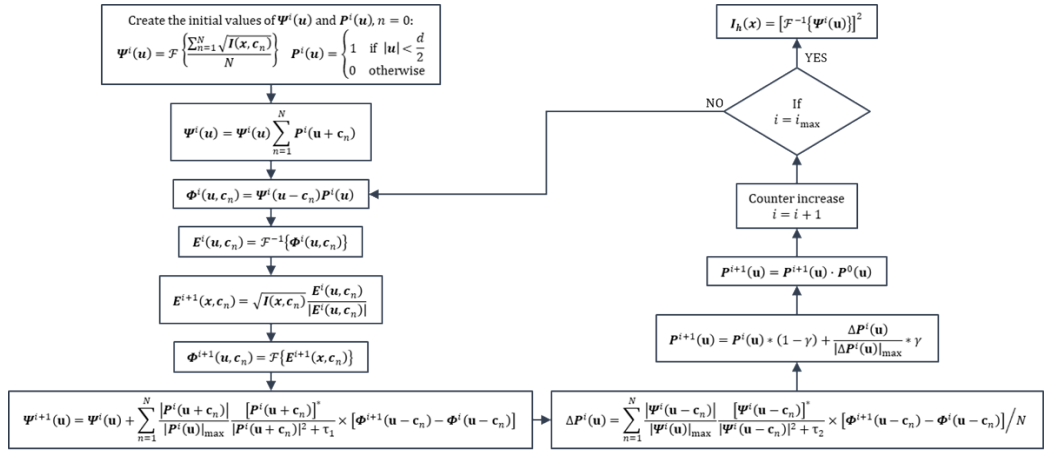

**Figure 5.** Flow chart of recovery algorithm.

**Funding**

This research was supported by the National Natural Science Foundation of China (62222513, 62305345) and the National Key Research and Development Program of China (2021YFA1401003)

**Disclosures**

The authors declare no conflicts of interest.

**Data availability**

Data underlying the results presented in this paper are not publicly available at this time but may be obtained from the authors upon reasonable request.

**Supplemental document**

See Supplement 1 for supporting content.